\documentclass{PoS}

\title{The Baikal-GVD detector calibrations}

\ShortTitle{The Baikal-GVD detector calibrations}

\author{A.D.~Avrorin$^{,a}$, A.V.~Avrorin$^a$, V.M.~Aynutdinov$^a$, R.~Bannash$^g$, I.A~Belolaptikov$^b$, V.B.~Brudanin$^b$, N.M.~Budnev$^c$, G.V.~Domogatsky$^a$, A.A.~Doroshenko$^a$, R.~Dvornick\'y$^{b,h}$, A.N.~Dyachok$^c$, Zh.-A.M.~Dzhilkibaev$^a$, \speaker{L. Fajt}$^{h,i}$, S.V~Fialkovsky$^e$, A.R.~Gafarov$^c$, K.V.~Golubkov$^a$, N.S.~Gorshkov$^b$, T.I.~Gress$^c$, R.~Ivanov$^b$, K.G.~Kebkal$^g$, O.G.~Kebkal$^g$, E.V.~Khramov$^b$ , M.M.~Kolbin$^b$, K.V.~Konischev$^b$, A.V.~Korobchenko$^b$, A.P.~Koshechkin$^a$, A.V.~Kozhin$^d$, M.V.~ Kruglov$^b$, M.K.~Kryukov$^a$, V.F.~Kulepov$^e$, M.B.~Milenin$^a$, R.A.~Mirgazov$^c$, V.~Nazari$^b$, \fbox{A.I.~Panfilov$^a$}, D.P.~Petukhov$^a$ E.N.~Pliskovsky$^b$, M.I.~Rozanov$^f$, E.V.~Rjabov$^c$, V.D.~ Rushay$^b$, G.B.~Safronov$^b$, B.A.~Shaybonov$^b$, M.D.~Shelepov$^a$, F.~\u{S}imkovic$^{b,h,i}$, A.V.~Skurikhin$^d$, A.G.~Solovjev$^b$, M.N.~ Sorokovikov$^b$, I.~\u{S}tekl$^i$, O.V.~Suvorova$^a$, E.O.~Sushenok$^b$, V.A.~Tabolenko$^c$, B.A.~Tarashansky$^c$, S.A.~Yakovlev$^g$\\
$^a$ Institute for Nuclear Research, Russian Academy of Sciences, Moscow, 117312 Russia\\
$^b$ Joint Institute for Nuclear Research, Dubna, 141980 Russia\\
$^c$ Irkutsk State University, Irkutsk, 664003 Russia\\
$^d$ Institute of Nuclear Physics, Moscow State University, Moscow, 119991 Russia\\
$^e$ Nizhni Novgorod State Technical University, Nizhni Novgorod, 603950 Russia\\
$^f$ St. Petersburg State Marine Technical University, St. Petersburg, 190008 Russia\\
$^g$ EvoLogics Gmbh, Germany\\ 
$^h$ Comenius University, Mlynska Dolina F1, Bratislava, 842 48 Slovakia\\
$^i$ Czech Technical University in Prague, Prague, 128 00 Czech Republic\\
E-mail: \email{lukas.fajt@utef.cvut.cz}
}

\abstract{In April 2019, the Baikal-GVD collaboration finished the installation of the fourth and fifth clusters of the neutrino telescope Baikal-GVD. Momentarily, 1440 Optical Modules (OM) are installed in the largest and deepest freshwater lake in the world, Lake Baikal, instrumenting 0.25 km$^3$ of sensitive volume. The Baikal-GVD is thus the largest neutrino telescope on the Northern Hemisphere. The first phase of the detector construction is going to be finished in 2021 with 9 clusters, 2592 OMs in total, however the already installed clusters are stand-alone units which are independently operational and taking data from their commissioning.

Huge number of channels as well as strict requirements for the precision of the time and charge calibration (ns, p.e.) make calibration procedures vital and very complex tasks. The inter cluster time calibration is performed with numerous calibration systems. The charge calibration is carried out with a Single Photo-Electron peak. The various data acquired during the last three years in regular and special calibration runs validate successful performance of the calibration systems and of the developed calibration techniques. The precision of the charge calibration has been improved and the time dependence of the obtained calibration parameters have been cross-checked. The multiple calibration sources verified a 1.5 - 2.0\,ns precision of the in-situ time calibrations. The time walk effect has been studied in detail with in situ specialized calibration runs.
}

\FullConference{36th International Cosmic Ray Conference -ICRC2019-\\
		July 24th - August 1st, 2019\\
		Madison, WI, U.S.A.}

\begin{document}

\section{Introduction}

The installation of the neutrino telescope Baikal-GVD started in 2015 and recently, more than half of its final sensitive volume is already instrumented by 1440 light sensitive Optical Modules (OMs). Their main goal, in the darkness in depth of 1 km of Lake Baikal, is the detection of Cherenkov light produced by the ultra relativistic secondary charged particles created in the interactions of high energy neutrinos. 


With advanced processing techniques the individual hits of optical modules can be used to reconstruct the trajectory, position and also the energy of the secondary particle and thus also the properties of the incident neutrino. The neutrinos interact very sporadically and due to their zero electric charge they are not affected by the galactic and extragalactic magnetic fields. This makes them ideal cosmic probes since the reconstructed direction of arrival points back to the place of their origin. Nevertheless, the neutrino detection and also the reconstruction procedure are very complex tasks complicated even more by the noise of the PMTs, lake and irreducible background in the form of atmospheric muons. For the best performance and highest accuracy of the detector, the calibration procedures have to be thoroughly adjusted and improved. 


\section{Time Calibration With LED Matrices}

The OMs in every cluster are divided to 24 sections (12 OMs/section). Signals from all OMs in a section are processed by 12-bit, 12 channel, 200\,MHz FADC located in the Central Module (CeM). The relative intra-section time calibration between OMs in the same section using the variety of independent calibration systems have been already developed, automatized and tested with obtained precision about 2\,ns \cite{icrc2017}. Moreover, the studies of the time variations of the intra-section calibration corrections have verified the satisfying level of stability ($\sigma = 0.27$\,ns) during more than 6 months.  

To extend the time calibration procedure to the whole cluster, the inter-section calibration step had to be implemented. For this task several LED matrices installed throughout every cluster have been used. An LED matrix is the 17" pressure resistant glass sphere which encompasses 12 Kingbright L-7113PBC-A LEDs. Six of them are oriented horizontally, in 3 different azimuths, and six of them vertically.  The LED matrix intensity is adjustable in the 1 - 10$^{9}$ photons range per flash. The LEDs flashing frequency can be set
from 1 up to 10$^4$\,Hz. Besides if necessary, the time delay between vertical and horizontal LEDs can be applied in the range 0 - 10$^3$\,ns. The real positions of the LED matrices and OMs are measured with the acoustic positioning system \cite{positioning}.

A brand new automatic inter-section time calibration module has been written. This module processes the regular LED matrix calibration runs performed regularly every week. The module automatically divides the calibration runs to the sub-runs, identifies the operation LED matrix in the sub-run, and reads and sets the real positions of all OMs and matrices. For every OM with enough hits, the expected time of the light detection from given LED matrix is calculated based on its distance and this value is compared with the real measured values. The obtained distribution of the time residuals (measured time - expected time) are fitted in the two step fitting procedure with the Gaussians. The two step procedure is applied to reduce the effect of the long tale in the residuals distribution caused by the scattering. The deviations of Gaussian mean $\mu$ from zero are interpreted as the time calibration corrections. 

\begin{figure}[h!]
  \centering
    \includegraphics[width=0.8\textwidth]{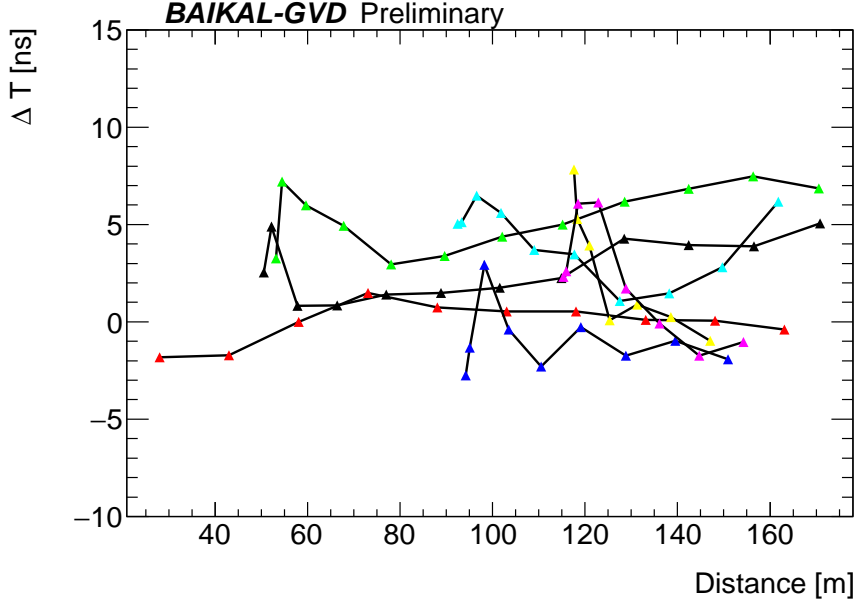}
  \caption{Results of the inter-section calibration from one of the sub-runs of the LED matrix calibration run. The figure shows the time residual versus the distance of the individual OMs from the LED matrix.}
  \label{fig:ledMatrix}
\end{figure} 

The results of the processing can be used to cross-check the precision of the previous intra-section calibration as well as to produce inter-section calibration corrections. Since the intra-section calibration corrections are applied to the LED matrix runs in the beginning of the calibration procedure, all the OMs in the given section should give the same inter-section offset caused by the differences in the sections cables and electronics. The variation $\sigma$ between OMs in a section gives the precision of the intra-section calibration. The mean value of the time residuals $\bar{T}_c$ from all OMs in a section gives the section time calibration correction.

The results of one calibration sub-run with a single LED matrix are shown in Fig. \ref{fig:ledMatrix}. The figure shows the time residuals $\Delta T$ of the individual OMs with respect to their distance from the LED matrix. The OMs are divided into groups representing individual sections. The variations between OMs in the individual sections present the precision of the intra-section calibration step. The results are summarized in Tab. \ref{tab:interSectionRes}. The table shows the number of OMs illuminated by the LED in sections, the average horizontal distance of the section from the LED matrix $\bar{d}_{matrix}$, average time residual of the section $\bar{T}_c$ used for the inter-section calibration, and precision of the intra-section calibration $\sigma$.

\begin{table}[ht]
\centering
\begin{tabular}{ c | c | c | c | c | c }
  \hline      
  String & Section & No. hit OMs & $\bar{d}_{matrix}$ [m] & $\bar{T}_c$ [ns] & $\sigma$ [ns] \\
  \hline
  \hline
  1 & upper & 12 & 50.57 &  2.76  & 1.57 \\
  2 & upper & 10 & 0     &  -0.05 &  1.04 \\
  3 & upper & 12 & 53.16 &  5.37  & 1.61 \\
  4 & upper &  9 & 94.18 &  -0.97 &  1.68 \\
  5 & upper &  7 & 117.65 & 2.47  & 3.26 \\
  6 & upper &  8 & 115.33 & 1.99  & 2.97 \\
  7 & upper & 10 & 92.49  & 4.09  & 1.90 \\
  \hline
\end{tabular} 
\caption{Results of the LED matrix inter-section calibration. Detailed description is given in the text.}
\label{tab:interSectionRes}
\end{table}

\section{Time Walk Effect (TWE)}

The TWE is an effect influencing the exact time of pulse detection according to its charge or amplitude. The pulses with higher amplitude start sooner and thus cross the threshold level earlier than the pulses with smaller amplitude. This can introduce the additional charge/amplitude dependent delay. The strength of this effect is dependent firstly on the shape of the pulse and secondly on the method of the pulse time extraction. To reach a nanosecond precision in the pulse time determination in a very wide charge range, the influence of TWE on the pulse time estimation has to be studied and the Time Walk Correction (TWC) function has to be used to compensate this effect.

\begin{figure}[h!]
  \centering
    \includegraphics[width=0.8\textwidth]{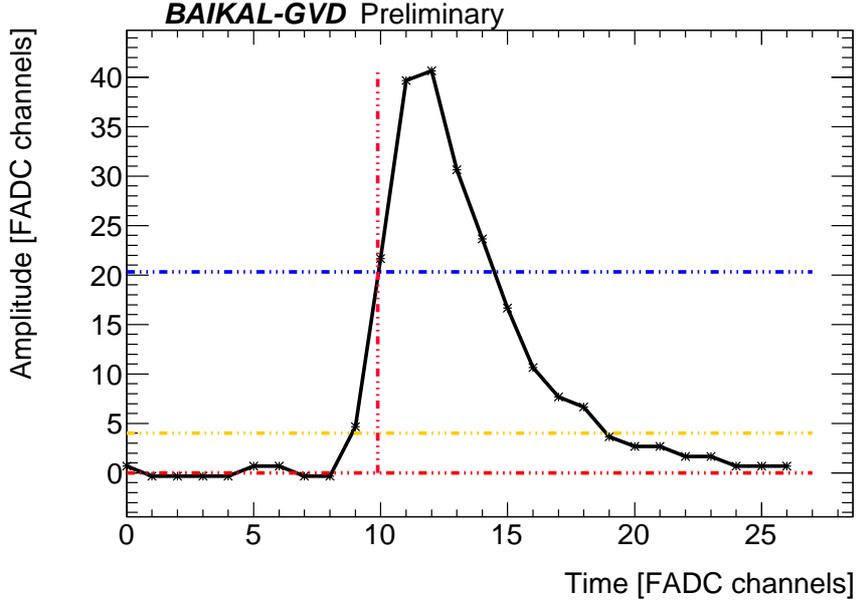}
  \caption{Illustration of the pulse time extraction procedure. Red and yellow horizontal lines represent pedestal and 4$\sigma$ threshold levels, respectively. Blue horizontal line marks the half of the pulse amplitude. The extracted pulse time is illustrated with red vertical line.}
  \label{fig:timeExtraction}
\end{figure} 

In the Baikal-GVD, time of the pulse is determined at the half of the maximal amplitude (see Fig. \ref{fig:timeExtraction}). With this time extraction method the strength of the TWE is reduced however not eliminated completely. The pulses with higher charge ($> 80$\,p.e.) saturate the FADC amplitude range however the processing circuit was designed to preserve measured charge. It means that TWC function is measured with respect to the charge and not with respect to the amplitude.

Until now, the TWE was studied only in the laboratory conditions for a few OMs and it would be very time consuming to measure it for all modules. To obtain the TWC function for all OMs in shorter time and in natural medium, the new extended calibration mode measured \emph{in situ} was proposed. The extended calibration run consists of more sub-runs. In each sub-run the built-in LEDs produce light of different intensity. Therefore, the pulses of different charge are produced and change in their detection time in different sub-runs can be studied.

By default, the 18 sub-runs are measured within one extended calibration run. The values of LED intensities were chosen to uniformly cover a charge range (0-700)\,p.e.. In every sub-run, the PMT transit time of all OMs is measured with test pulses and built-in LEDs. According to the intensity of the LED, the pulses of different charge are produced and used for the transit time calculation. Therefore, the change in the PMT transit time between individual sub-runs is given by the TWE. In every sub-run the average charge of the pulses used for the transit time measurement is calculated. Thus, the dependency of the transit time on the average charge can be constructed. 

\begin{figure}[h]
\begin{center}
  \includegraphics[width=0.8\linewidth]{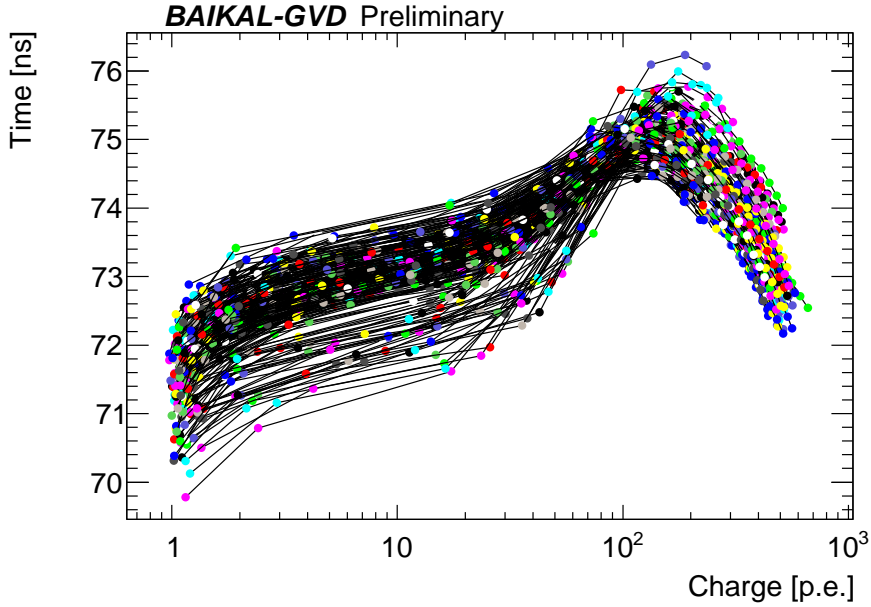}
\caption{The TWE measured in situ with 18 different intensities of the calibration LEDs.}
\label{fig:tweResults}
\end{center}
\end{figure}

The results of TWE measurement for all operating OMs from one of the clusters are depicted in the Fig. \ref{fig:tweResults}. The slight vertical shifts of the whole curves representing individual OMs are present. However, the overall behavior of the curves is similar. From this distribution the TWC function can be constructed. As can be seen from the figure, the TWE reaches up to 4\,ns difference in the studied region. This value significantly exceeds the timing precision of the calibration procedure and thus has to be taken into account especially for the events with large difference in deposited charge in the OMs (cascade like events, LED matrix calibration runs, etc.).

\begin{figure}[h]
\begin{center}
  \includegraphics[width=0.8\linewidth]{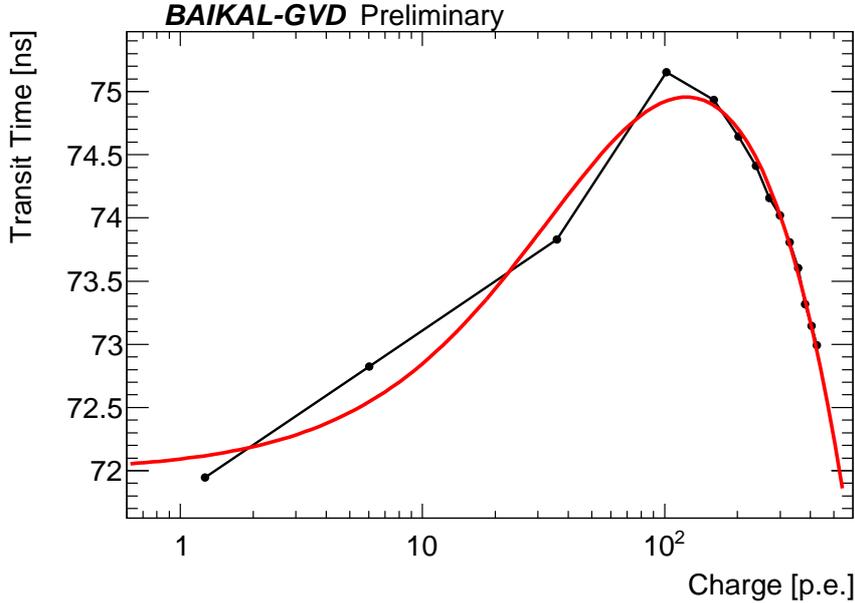}
\caption{An example of the fit of the TWE measurements. The result of the fit is used as a TWC function of the given OM.}
\label{fig:tweMeas}
\end{center}
\end{figure}

The TWE measured for every individual OM is fitted by a particular TWC function: 

\begin{equation}
  f(Q) = a - \frac{b}{Q-c} + d\,Q \ ,
\end{equation}

\noindent where $Q$ represents charge in photoelectrons, $a$, $b$, $c$, and $d$ are free parameters. The result of the fit for one individual OM is shown in Fig. \ref{fig:tweMeas}. The deviations of the measured and fitted values are safely below 0.5 ns. The obtained parameters are saved in the database and can be used during data processing.

\section{Charge Calibration}

The new algorithm for the pulse charge calculation has been introduced to the data processing. In the previous algorithm, the charge was integrated only in the samples with values higher than a threshold. A threshold is calculated based on the variations of the pedestal voltage from the first few samples in the waveform which precede the pulse itself. The threshold is equal to $4\,\sigma$ of the pedestal variations. If the $\sigma$ is smaller than 1 FADC channel it is automatically set to 1 what means that the smallest possible threshold is 4 FADC channels.

\begin{figure}[h!]
\begin{center}
	\includegraphics[width=0.8\linewidth]{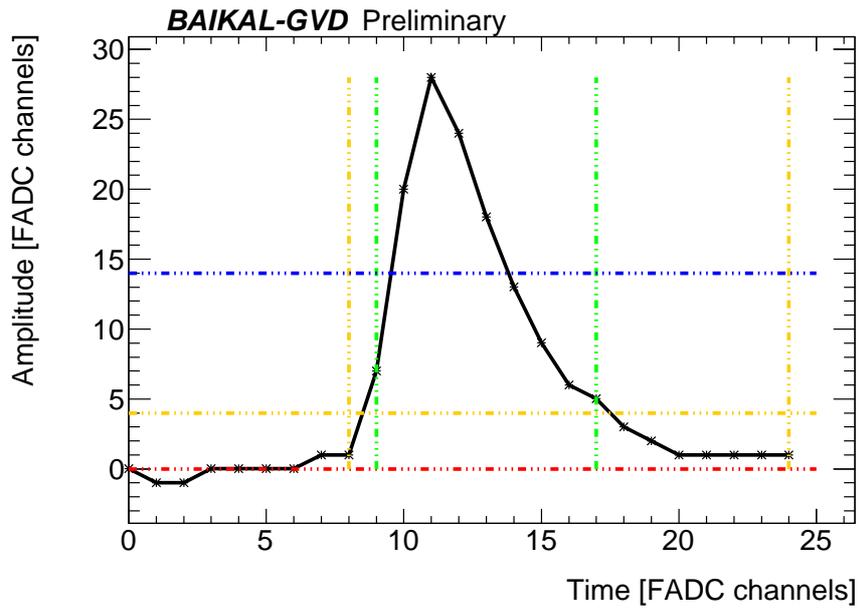} 
\caption{The illustration of the previous and new pulse extraction techniques. The pedestal is drawn by red horizontal line, the $4\,\sigma$ threshold is shown by yellow horizontal line and the half amplitude is illustrated by blue horizontal line. The green and yellow vertical lines show previous and new pulse charge integration region, respectively. }
\label{fig:newProcessingExamples}
\end{center}
\end{figure}

The new extraction technique uses $4\,\sigma$ threshold as well to detect a presence of the pulse, however the charge integration is extended even on samples below the threshold at the beginning and at the end of the pulse. This extension causes approximately 10-15\,\% increase in the integrated charge of the small pulses (1-2\,p.e.) whose charge is used for the charge calibration of the OMs. The comparison of the previous and new extraction techniques is shown in Fig. \ref{fig:newProcessingExamples}. 

\begin{figure}[h!]
\begin{center}
	\includegraphics[width=0.8\linewidth]{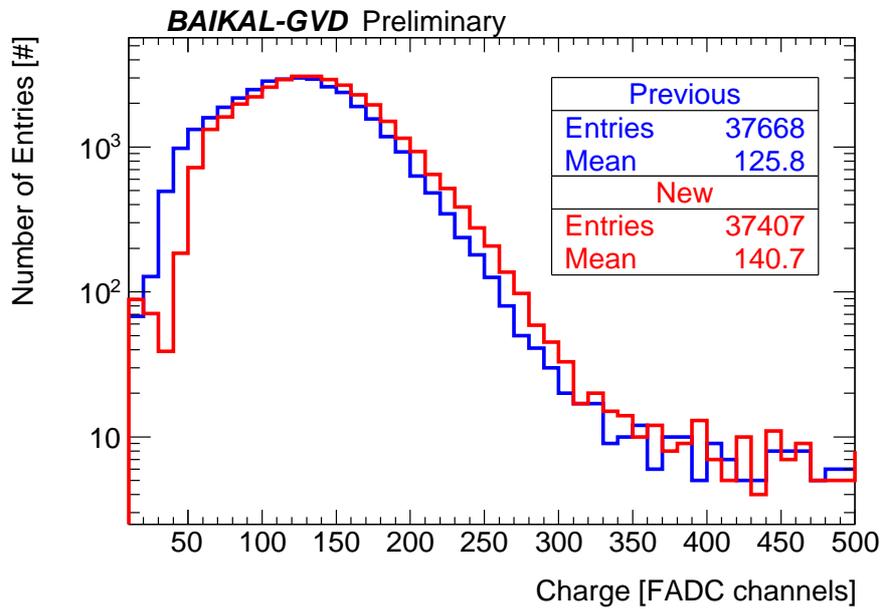} 
\caption{Comparison of the charge distribution of the single photoelectron peak with the previous (blue) and new (red) pulse extraction technique. The increase in the extracted charge is 11.8\,\%.}
\label{fig:chargeDistComp}
\end{center}
\end{figure}

The new extraction will influence mainly a small pulses since the relative charge increase is large with respect to the overall pulse charge. At the same time, the single photoelectron peak is used for the charge calibration of the OMs, i.e. the mean value of the charge (in FADC channels) of the single photoelectron peak is assign 1\,p.e.. With the new pulse extraction technique the mean value increases about 11.8\,\% (see Fig. \ref{fig:chargeDistComp}).

\begin{figure}[h!]
\begin{center}
	\includegraphics[width=0.45\linewidth]{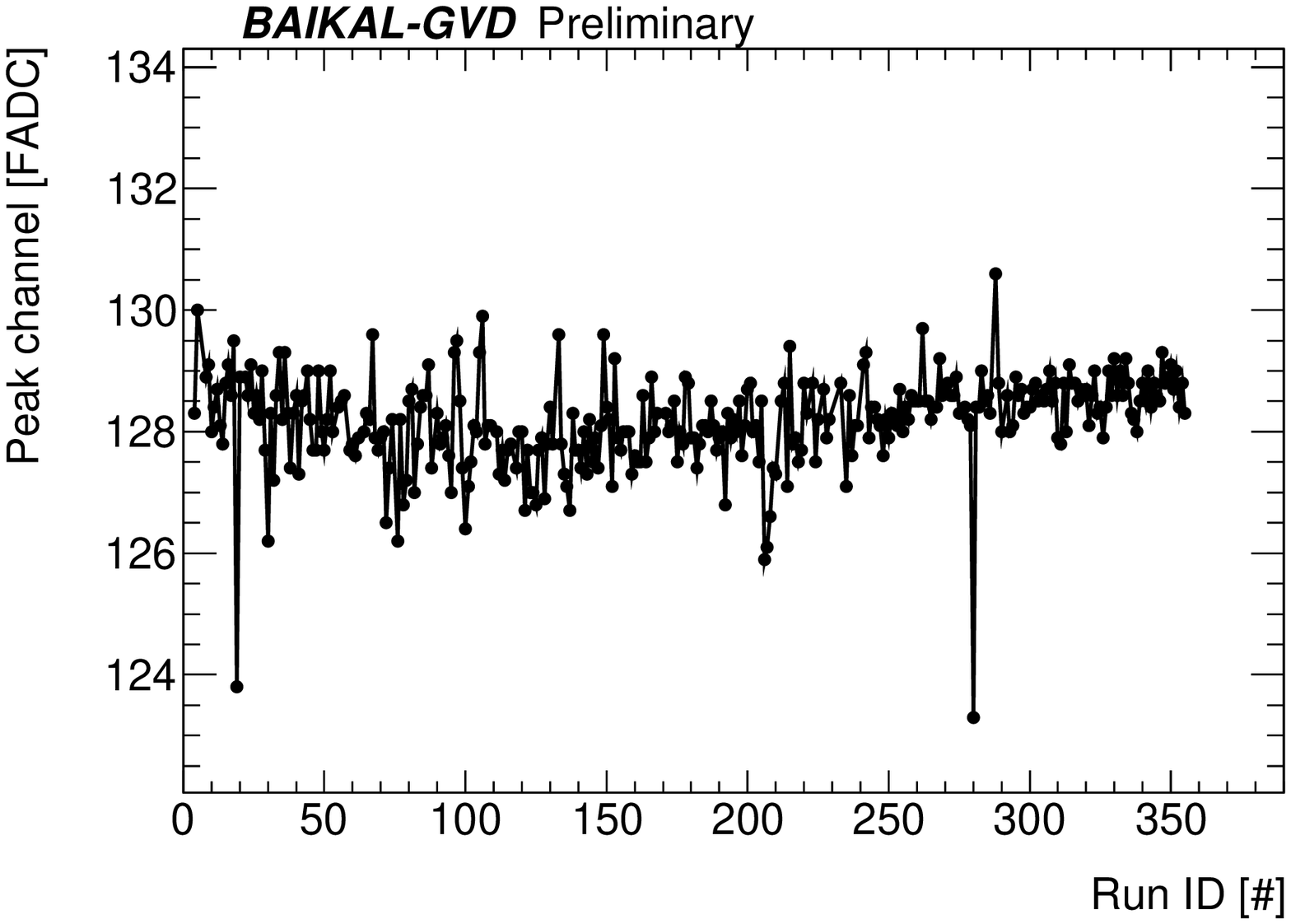} 
	\includegraphics[width=0.45\linewidth]{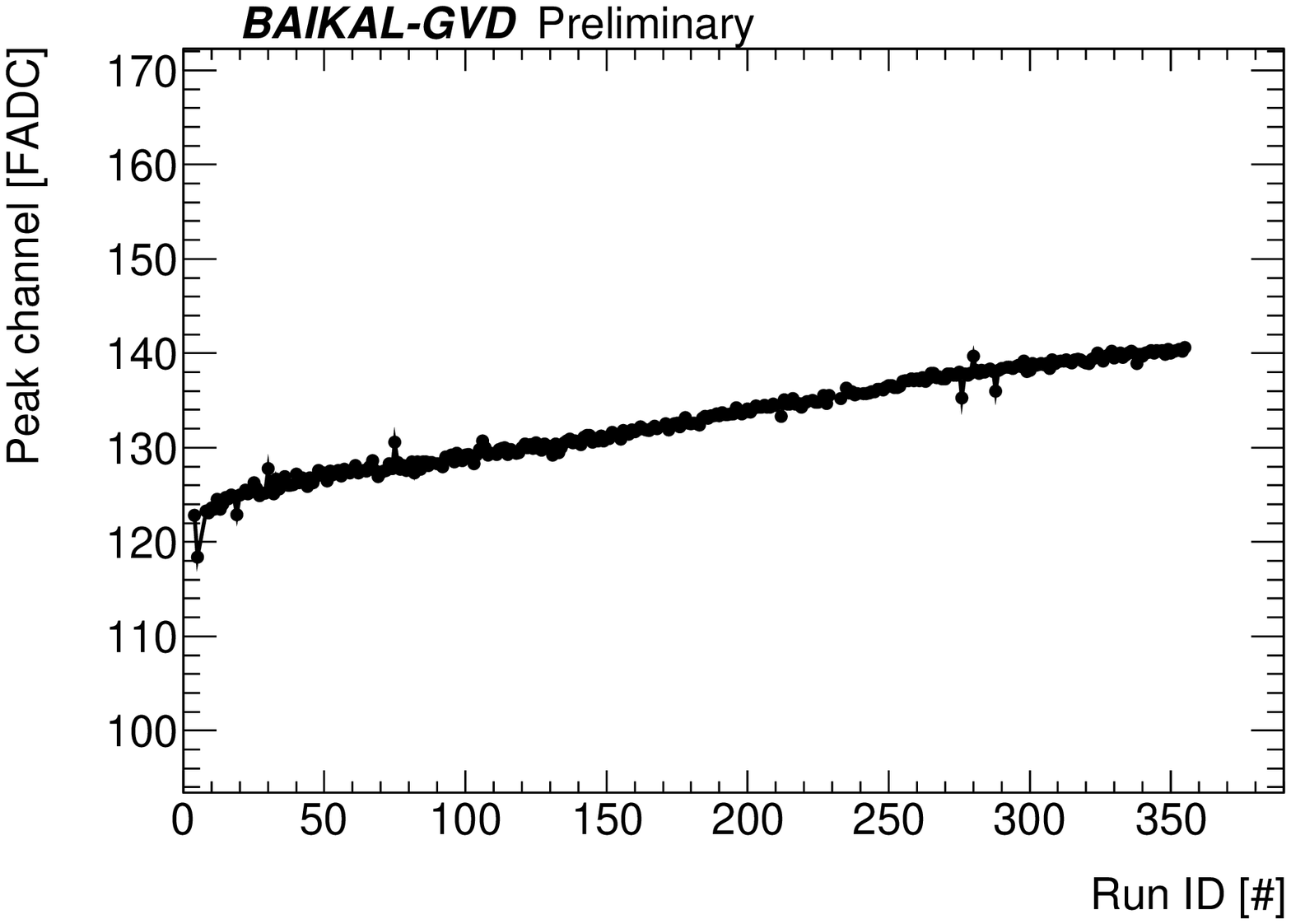} 
\caption{Examples of the time stability of the charge calibration constants from two OMs. The right figure shows the significant increase in the value of the calibration constant which has not been explained yet.}
\label{fig:qCalTimeStab}
\end{center}
\end{figure}

The position of the single photoelectron peak is stored in the database for every channel for every run. These charge calibration constants are used to transform FADC channels to photoelectrons. The time stability of the calibration constants have been studied. The examples for two random channels are shown in Fig. \ref{fig:qCalTimeStab}. The majority of the OMs reports very stable behavior (1-2\%) during the year. However, on approximately 10\,\% OMs a significant increase of the value of the calibration constant has been observed (right figure). This behavior has not been explained yet however the change in the pulse shape in the course of year is being studied in detail.

\vspace{-0.5cm}

\section{Conclusion}

The brand new automated processing of the LED matrix runs have been used for the inter-section calibration of the operating clusters. Moreover, the new processing offers the cross-check of the intra-section time calibration precision obtained with other calibration light sources. The achieved precision is significantly below 2\,ns for OMs closer to 100 meters. 

The strength of the Time Walk Effect on the pulse time extraction precision in situ has been evaluated. In the worst case scenario, the TWE can reach up to 4\,ns depending on the charge of the pulse and thus has to be undoubtedly taken into account. To eliminate this effect, two different approaches were used to design the correction function.

The new algorithm for the pulse extraction region was implemented which led to the increase of the charge of the small pulses (a few photoelectrons) about 10-15 \%. Since the charge calibration of the OMs is based on the charge of the single photoelectron pulse, the new pulse processing technique improves the precision of the charge calibration of the whole detector.

\section{Acknowledgements}

This work was supported by the Russian Foundation for Basic Research (Grants 16-29-13032, 17-0201237 and by the Ministry of Education, Youth and Sports of the Czech Republic under the contract number CZ.02.1.01/0.0/0.0/16\_013/0001733.

\end{document}